\documentclass{article}
\usepackage{fdl,spconf}
\title{A Splitting-Based Iterative Algorithm for GPU-Accelerated Statistical Dual-Energy X-Ray CT Reconstruction}
\name{Fangda Li \qquad Ankit Manerikar \qquad Tanmay Prakash \qquad Avinash Kak \thanks{This research is funded by the BAA 17-03 AATR contract with Department of Homeland Security.}} 
\address{School of Electrical and Computer Engineering, Purdue University \\
\texttt{\{li1208, amanerik, tprakash, kak\}@purdue.edu}
}

\begin{document}
%\ninept
\maketitle
\begin{abstract}

When dealing with material classification in baggage at
airports, Dual-Energy Computed Tomography (DECT) allows
characterization of any given material with coefficients
based on two attenuative effects: Compton scattering and
photoelectric absorption.  However, straightforward
projection-domain decomposition methods for this
characterization often yield poor reconstructions due to the
high dynamic range of material properties encountered in an
actual luggage scan.  Hence, for better reconstruction
quality under a timing constraint, we propose a
splitting-based, GPU-accelerated, statistical DECT
reconstruction algorithm.  Compared to prior art, our main
contribution lies in the significant acceleration made
possible by separating reconstruction and decomposition
within an ADMM framework.  Experimental results, on both
synthetic and real-world baggage phantoms, demonstrate a
significant reduction in time required for convergence.

\end{abstract}

\vspace{-0.12in}
\section{Introduction}
\vspace{-0.12in}
\label{sec:intro}
X-ray Computed Tomography (CT) is a widely deployed
technique for threat detection in baggage at airport
security checkpoints.  However, using a single X-ray
spectrum limits its reconstruction to only that of LAC
(linear attenuation coefficient) or HU (Hounsfield Unit)
images which are, at best, an approximation to the underlying
energy-dependent characteristics of the materials.  While
LAC might suffice for discriminating between different types
of tissues in medical imaging, in the adversarial case of
airport baggage scanning, materials used commonly for
homemade explosives can possess LAC values that are nearly
the same as for benign materials.  As a result, the
technique of DECT, where projections from two X-ray spectra
are collected simultaneously at each angle, has been
proposed for material discrimination because it allows for
the material property to be recovered in an additional
dimension by using an energy-dependent attenuation model.

The most commonly used DECT model represents X-ray
attenuation as the combined effect of Compton scattering and
photoelectric absorption, which can be written as:
% \vspace{-0.12in}
\begin{equation}
% \vspace{-0.06in}
\mu(E, x) = x_c f_{\text{KN}}(E) + x_p f_p(E), 
\label{eq:attenuation}
\end{equation}
where $f_{\text{KN}}(E)$ and $f_p(E)$ denote the
energy-dependent multipliers\footnote{See
  \cite{ying2006dual} for a detailed formulation of
  $f_{\text{KN}}(E)$ and $f_p(E)$.} to the Compton
coefficient $x_c$ and the photoelectric (PE) coefficient
$x_p$, respectively.  Therefore, the goal of DECT is to
recover both the Compton and the PE coefficients of the
object using the projections $\vm_h, \vm_l \in \R^M$
measured at two different X-ray spectra.  This amounts to
solving simultaneously for $\vx_c, \vx_p \in \R^N$ from
the following two equations that can be shown compactly 
by:
% \vspace{-0.06in}
\begin{equation}
% \vspace{-0.12in}
\vm_{h/l} = - \ln \int S_{h/l}(E) e^{-\mR \vmu(E)} dE + \ln \int S_{h/l}(E) dE,
\end{equation}
where $\mR$ denotes the forward projection matrix, and
$S_{h}(E)$ and $S_{l}(E)$ denote, respectively, the photon
distribution across energy levels in the high- or low-energy
spectra.

Solving the two equations shown above is made challenging by
the fact the two phenomena that contribute to X-ray
attenuation --- Compton and PE, occur at grossly
different scales.  At most applicable energy levels, Compton
scattering is the dominant contributor to attenuation and
this disparity between the two phenomena becomes worse with
increasing energy since $f_p(E)$ has a cubic decay with $E$.
As a result, stable recovery of PE coefficients requires
more sophisticated inversion algorithms, such as those
described in \cite{semerci2014tensor, zhang2017tensor,
  semerci2012parametric, tracey2015stabilizing}.

Unfortunately, the algorithms cited above tend to be
iterative and, with run-of-the-mill computing hardware, take
a long time to return the results.  Therefore, a
straightforward implementation of these algorithms is not
appropriate for the end-goals that motivate our research ---
high-throughput baggage screening at airports.  The focus of
the concepts presented in this paper is on \textit{improving the computational efficiency} of an ADMM-based statistical
inversion algorithm.

\vspace{-0.12in}
\section{Related Work}
\vspace{-0.12in}
\label{sec:related}
DECT involves two key steps: dual-energy decomposition and
tomographic reconstruction.  The two tasks can be done
either sequentially, as in projection-wise
decomposition, or in a unified step using
iterative statistical approaches.

\noindent\textbf{Projection-wise Decomposition}: One of the
earliest approaches for DECT decomposition, the
Constrained Decomposition Method (CDM) \cite{ying2006dual}
involves directly decomposing the dual energy projections to Compton and PE line integrals followed by the FBP reconstructions of the two. In \cite{yuan2016robust}, CDM was also extended to operate for Multi-Energy CT.  A major disadvantage of this method is that it
guards itself poorly against artifacts, especially in PE
coefficients but it is still a preferred approach as it 
enables parallel implementation.

\noindent\textbf{Iterative Statistical Approaches}: The
statistical methods for DECT solve for the MAP estimates, finding 
the Compton and PE coefficients that best correspond to
the measurements and any prior knowledge.  The literature on
Multi-Energy CT has focused largely on designing the models
and the priors that best leverage the structural similarity
across bases.  In \cite{semerci2012parametric}, Compton/PE
images are reconstructed on a set of exponential basis
functions with an edge-correlation penalty term.  This idea
of encouraging geometric similarity between the more stable
Compton image and the PE image is further explored in
\cite{tracey2015stabilizing}.  For this purpose, the authors have
proposed a new Non-Local Mean (NLM) regularizer on the PE
image and laid out an ADMM formation that scaled up to a
problem size of practical interest.  By treating the energy
level as a third dimension, the contributions in
\cite{semerci2014tensor, zhang2017tensor} adopt a
tensor-based model for reconstruction using sparse priors.

Despite being able to produce high-quality DECT, statistical
reconstructions generally fail drastically with regard to the
timing constraints of practical applications.
Compared to LAC
reconstructions, DECT has to deal with the added
computational burden associated with the decomposition step.
As a result, in existing approaches, solving decomposition
and reconstruction in one step becomes inefficient due to
the combined complexity and high-dimensionality of the
problem.

We therefore
propose a statistical DECT approach that combines the best
of the projection-wise decomposition methods and the
iterative statistical methods.  More specifically, we employ
a splitting-based MAP estimation method embedded in an ADMM
framework.  As we will show in Section \ref{sec:exp}, our new
splitting scheme not only provides a better convergence
rate, but also allows for powerful hardware-enabled
acceleration.

\vspace{-0.12in}
\section{Proposed Method}
\label{sec:method}
\vspace{-0.12in}
\subsection{Problem Formulation}
To describe our proposed method, we first define the forward
model for X-ray attenuation, i.e., the nonlinear transform
$f(\cdot)$ from Compton/PE coefficients to logarithmic
projections:
\begin{equation}
\begin{aligned}
f_{h/l}(\mRtilde \vx) = C_{h/l} - \ln \int S_{h/l} e^{- f_{\text{KN}} \mR \vx_c - f_{p} \mR \vx_p} dE
\end{aligned}
\end{equation}
where $C_{h/l}$ are constants, 
% $\vx = \begin{bmatrix} \vx_c \\ \vx_p \end{bmatrix}$ 
$\vx = \begin{bmatrix} \vx^T_c; \vx^T_p \end{bmatrix}^T$ 
  and $\mRtilde = \begin{bmatrix}
  \mR; \mR \end{bmatrix}$.  Given a pair of line-integral
measurements 
% $\vm = \begin{bmatrix} \vm_h \\ \vm_l \end{bmatrix}$
$\vm = \begin{bmatrix} \vm^T_h; \vm^T_l \end{bmatrix}^T$
corrupted by Poisson photon noise
and possibly other artifacts, we construct a MAP estimate
with the Total Variation (TV) term and the non-negativity
prior:
\begin{equation}
\begin{aligned}
\vxhat_{\text{MAP}} = \argmin{\vx} \frac{1}{2} \left\| f(\mRtilde \vx) - \vm \right\|^2_\Sigma + \lambda |\vx|_{\text{TV}} + g(\vx),
\end{aligned}
\label{eq:map}
\end{equation}
where $\Sigma$ is a diagonal matrix with photon counts as trace elements as proposed in \cite{bouman1996unified}, $g(x_i) = \left\{\begin{matrix}
0, x_i \in \R^+ \\ 
\infty, x_i \notin \R^+
\end{matrix}\right.$ and $\lambda$ is the regularization parameter.
While the unconstrained optimization in \eqref{eq:map} is
% both intractable and 
highly inefficient to solve directly, the
Alternating Direction Method of Multipliers (ADMM) method
provides a flexible splitting-based framework for
both optimality and convergence
\cite{tracey2015stabilizing}.

\vspace{-0.12in}
\subsection{Formulation for ADMM}
%% The derivation for solving for the MAP estimate using
ADMM
begins with the conversion of \eqref{eq:map} to its
constrained equivalent.  By introducing two new auxiliary
variables $\vy$ and $\vz$ for the TV and the non-negativity
terms, respectively, and by posing $\vx$ as the primal
variable, the MAP minimization can be transformed into:
\vspace{-0.06in}
\begin{align}
\vxhat_{\text{MAP}} = &\argmin{\vx, \vy, \vz}  \frac{1}{2} \left\| f(\mRtilde \vx) - \vm \right\|^2_\Sigma + \lambda |\vy| + g(\vz), \nonumber\\
&\text{s.t.} \begin{bmatrix} \vy_{c/p} \\ \vz_{c/p} \end{bmatrix} = \begin{bmatrix}
\mD \\ \mI \end{bmatrix}\vx_{c/p} = \mC \vx_{c/p},
\label{eq:map_constrained}
\end{align}
where $\mD$ denotes the finite difference operator.
Subsequently, the corresponding Augmented Lagrangian (AL)
can be written as:
\vspace{-0.12in}
\begin{equation}
\begin{aligned}
L(\vx, \vy, \vz) = &\frac{1}{2} \left\| f(\mRtilde \vx) - \vm \right\|^2_\Sigma + \lambda |\vy| + g(\vz) \\
&+ \sum_{\beta \in \{c, p\}} \frac{\rho_\beta}{2} \left\| \begin{bmatrix} \vy_\beta \\ \vz_\beta \end{bmatrix} - \mC \vx_\beta + \vu_\beta \right\|^2,
\end{aligned}
\label{eq:al}
\end{equation}
where $\vu$ denotes the dual variable, or the scaled
Lagrangian multiplier, and $\rho$ denotes the penalty
parameter.  ADMM splits the minimization of AL into
subproblems that are solved separately in an iterative
framework.  One such intuitive splitting scheme proposed in
\cite{tracey2015stabilizing} is for the iterative updates to
be carried out according to \eqref{eq:splitc} through
\eqref{eq:splitu} shown below.  First, the primal variables
are updated using:
\vspace{-0.06in}
\begin{equation}
\begin{aligned}
\vxhat_c = &\argmin{\vx_c} \frac{1}{2} \left\| f(\mR \vx_c, \mR \vx_p) - \vm \right\|^2_\Sigma \\
&+ \frac{\rho_c}{2} \left\| \begin{bmatrix} \vy_c \\ 
\vz_c \end{bmatrix} - \mC \vx_c + \vu_c \right\|^2,    
\end{aligned}
\label{eq:splitc}
\end{equation}
\vspace{-0.06in}
\begin{equation}
\begin{aligned}
\vxhat_p = &\argmin{\vx_p} \frac{1}{2} \left\| f(\mR \vxhat_c, \mR \vx_p) - \vm \right\|^2_\Sigma \\
&+ \frac{\rho_p}{2} \left\| \begin{bmatrix} \vy_p \\ 
\vz_p \end{bmatrix} - \mC \vx_p + \vu_p \right\|^2.    
\end{aligned}
\label{eq:splitp}
\end{equation}
Note that the update to $\vx_p$ is made subsequent to
$\vxhat_c$ since we can expect $\vxhat_c$ to stabilize the
recovery of the more noise-prone $\vx_p$.  
% Secondly, we update the auxiliary variables $\vy$ and $\vz$ corresponding to the TV and non-negativity terms, respectively, by using
Secondly, we update the auxiliary variables $\vy$ and $\vz$ corresponding to the TV and non-negativity terms, respectively, by using
\vspace{-0.12in}
\begin{align}
\vyhat_{c/p} &= \mathit{ shrinkage}(\mD \vxhat_{c/p} - \vu_{c/p}^y, \frac{\lambda_{c/p}}{\rho_{c/p}}),
\label{eq:splity} \\
\vzhat_{c/p} &= \max(0, \vxhat_{c/p} - \vu_{c/p}^z),
\label{eq:splitz}
\end{align}
where the shrinkage function is defined in 
% equation 9.35 in \cite{bouman2013model}. in 
\cite[Eq.\ 9.35]{bouman2013model}.
Then the dual variable $\vu$ is updated with
\vspace{-0.08in}
\begin{equation}
\begin{aligned}
\vuhat_{c/p} = \vu_{c/p} + \begin{bmatrix} \vyhat_{c/p} \\ \vzhat_{c/p} \end{bmatrix} - \mC \vxhat_{c/p}.
\end{aligned}
\label{eq:splitu}
\end{equation}
Lastly, we update $\rho$ adaptively using a method described
in \cite{boyd2011distributed} that is based on the primal
and dual residual.
%With regard to computational efficiency, while
While \eqref{eq:splity}-\eqref{eq:splitu} can be realized by
straightforward element-wise operations, solving
\eqref{eq:splitc}-\eqref{eq:splitp} requires a
nonlinear least squares algorithm such as
Levenberg-Marquardt (LM).  Despite its robustness, LM, is a
sequential algorithm, offering little room for
parallelization.
Intuitively, the overall computational inefficiency can be
attributed to \eqref{eq:splitc} or \eqref{eq:splitp}
which at once addresses two problems of very different nature:
nonlinear dual-energy decomposition and regular linear
tomographic reconstruction.

\begin{figure*}[t]
\centering
\includegraphics[width=0.87\textwidth]{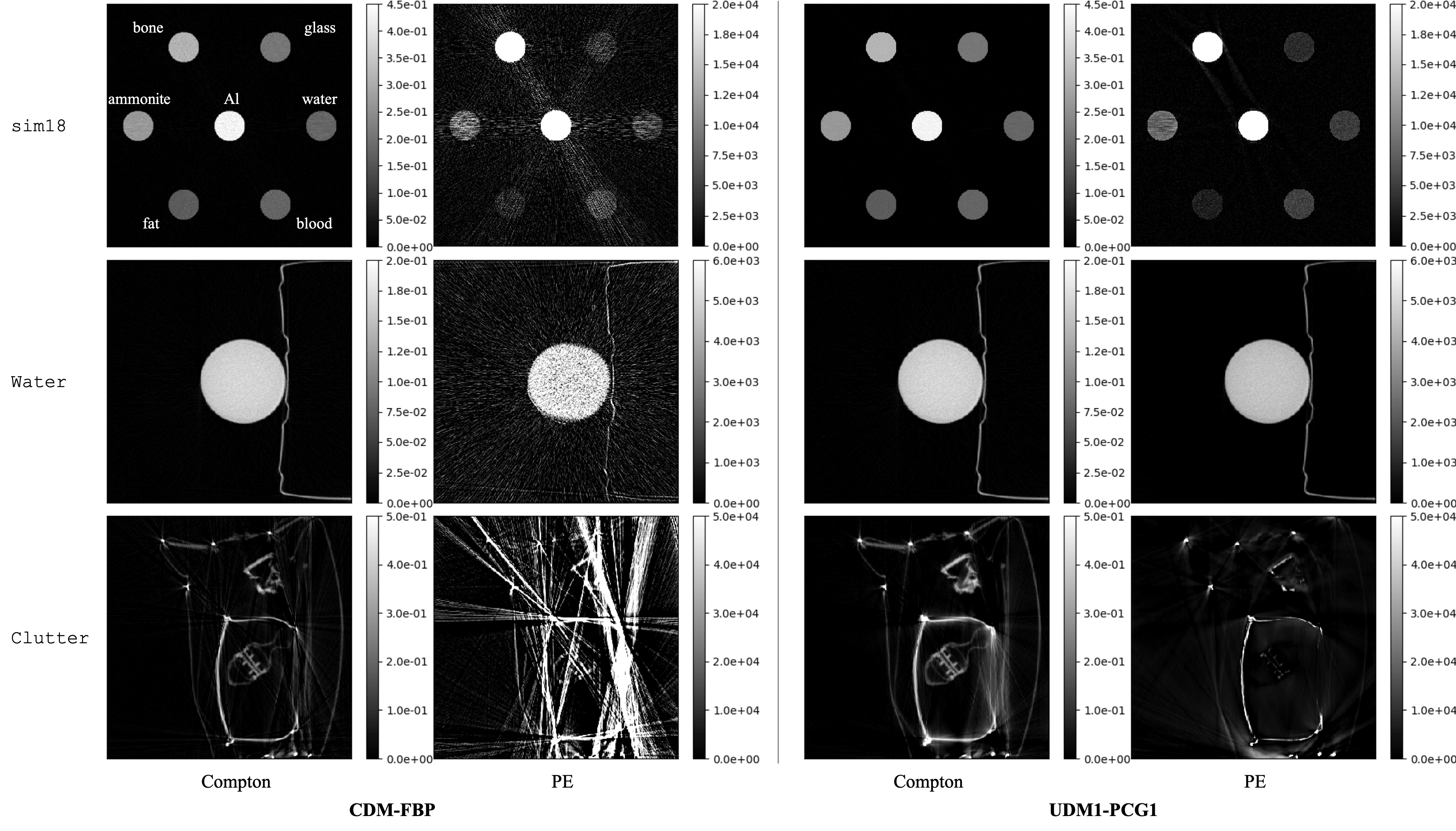}
\caption{ Compton and PE reconstructions by \textbf{CDM-FBP} and \textbf{UDM1-PCG1}. 
  Note that we clipped the values in PE images of \texttt{sim18} to show the streaking artifacts in the \textbf{CDM-FBP} reconstruction.
  Such artifacts, appearing also in the \textbf{CDM-FBP} reconstructions of \texttt{Water} and \texttt{Clutter}, are significantly
  suppressed by the \textbf{UDM1-PCG1} method.
}
\label{fig:recons}
\vspace{-0.12in}
\end{figure*}
\begin{figure}[h]
\centering
\includegraphics[width=0.44\textwidth]{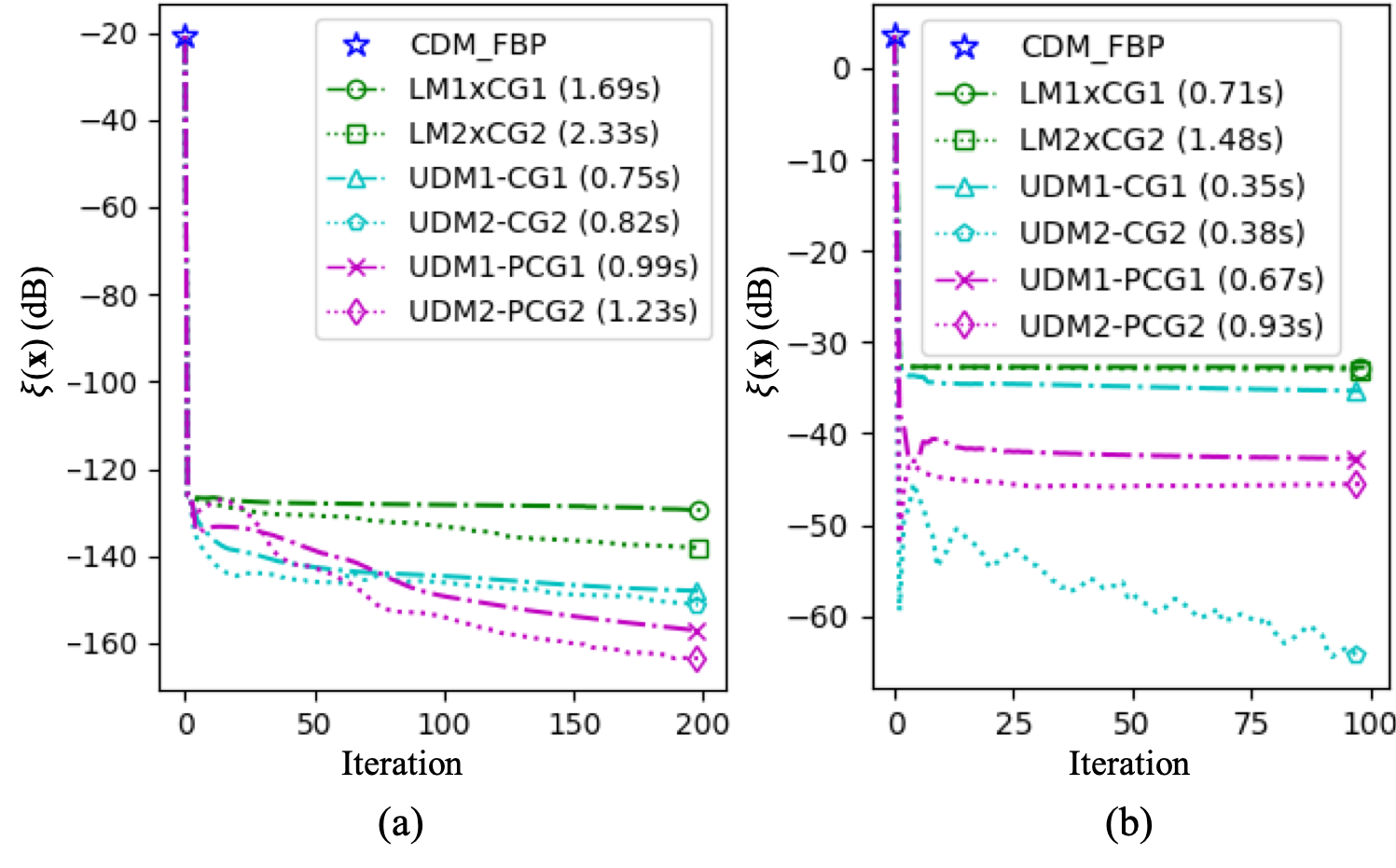}
\caption{
$\xi(\vx)$ vs.\ iteration for PE and average
seconds \textit{per iteration} for (a) \texttt{sim18} and
(b) \texttt{Water}.  The same plots for Compton are
omitted since they exhibit similar trends.
}
\label{fig:sim18_water_e_i}
\vspace{-0.12in}
\end{figure}

\vspace{-0.12in}
\subsection{Proposed Splitting Scheme}
For the new sped-up implementation presented in this
paper, we further decompose
\eqref{eq:splitc}-\eqref{eq:splitp} into two simpler
subproblems: tomographic reconstruction followed by
dual-energy decomposition. 
This is achieved by viewing measurement fitting as an additional constraint: $\va = \mRtilde \vx$, where $\va$ is a new auxiliary variable  for that purpose.
%% Using $\va$, for the first subproblem we pose tomographic
Incorporating this constraint with the others, for the first subproblem we pose tomographic
reconstruction as an unweighted unconstrained least squares
problem:
\vspace{-0.14in}
\begin{equation}
\vxhat_{c/p} = \argmin{\vx_{c/p}} \frac{\rho_{c/p}}{2} \left\| \begin{bmatrix} \va_{c/p} \\ \vy_{c/p} \\ \vz_{c/p} \end{bmatrix} - \begin{bmatrix} \mR \\ \mC \end{bmatrix} \vx_{c/p} + \begin{bmatrix} \vu^a_{c/p} \\ \vu^y_{c/p} \\ \vu^z_{c/p} \end{bmatrix} \right\|^2.
\label{eq:splitcg}
\end{equation}
and we can now separately deal with the decomposition, i.e.
solving for $\va$, in the following subproblem.
We have chosen to use a CG solver for
the minimization shown above because our system is huge and
sparse.  Additionally, since the projection-wise weights
$\Sigma$ are now absorbed by the decomposition step to be
described later, the linear system corresponding to
\eqref{eq:splitcg} is shift-invariant.  Therefore, for
improved convergence rate, each CG iteration lends itself
well to preconditioning using a high-passing filter \cite{Fessler1999Conjugate}.
In our implementation, we experimented with the ramp filter as the preconditioner.

% Implementation details of the FFT-based high-passing filter
% can be found in \cite{Fessler1999Conjugate}. 
% The acronym PCG is commonly used to denote solution strategies based on preconditioned conjugate gradient methods.

The second subproblem, that of dual-energy decomposition,
can be solved by finding the pair of Compton/PE coefficients
that minimizes the cost function at each ray projection
independently.  Therefore, the decomposition is achieved by
using the Unconstrained Decomposition Method (UDM):
\vspace{-0.10in}
\begin{equation}
\begin{aligned}
\begin{bmatrix} \vahat_c \\ \vahat_p \end{bmatrix} = &\argmin{\va_c, \va_p} \frac{1}{2} \left\| f(\begin{bmatrix} \va_c \\ \va_p \end{bmatrix}) - \vm \right\|^2_\Sigma \\
&+ \sum_{\beta \in \{c, p\}}\frac{\rho_\beta}{2} \left\| \va_\beta - \mR \vxhat_\beta + \vu_\beta^a \right\|^2
\end{aligned}
\label{eq:splitcdm}
\end{equation}
Compared to \eqref{eq:splitc} and \eqref{eq:splitp}, the
update formulas in \eqref{eq:splitcg} and
\eqref{eq:splitcdm} have the following advantages: First, treating
reconstruction and decomposition separately allows us to use
two entirely different algorithms, with each tailored for
the corresponding subproblem, which in our case are Preconditioned CG (PCG) and UDM. 
% Secondly, projection-wise decomposition is well-suited
% for large-scale parallelization.  For solving instances of
% \eqref{eq:splitcdm} simultaneously, we can now leverage
% powerful hardware specialized for mass parallelization such
% as GPUs.
As an additional benefit, while LM is still used in UDM, backtracking of the damping parameter, which is necessary for dealing with nonlinear optimization, no longer involves expensive tomographic operations.
Lastly, computational efficiency is improved in terms of both the number of operations required, as listed in Table \ref{tb:computation}, and parallelization.
We can now not only independently execute \eqref{eq:splitcg} for both bases, but also leverage massively parallelized hardware such as GPUs to solve \eqref{eq:splitcdm}.

\begin{table}[t]
\centering
\caption{Minimum number of operations per ADMM iteration. 
In \textbf{LM($m$)$\times$CG($n$)}, $n$ CG iterations are taken to compute the update per LM iteration. Note that LM may require more operations than listed to tune the damping parameter.
}
\begin{tabular}{@{}l|ccc@{}}
\toprule
 & $\mR$ & $\mR^T$ & $f(\cdot)$ \\ \midrule
\textbf{LM($m$)$\times$CG($n$)} & $2m(n+1)$ & $2m(n+1)$ & $4m$ \\
\textbf{UDM($m$)-PCG($n$)} & $2n$ & $2(n+1)$ & $4m$ \\ \bottomrule
\end{tabular}
\label{tb:computation}
\vspace{-0.12in}
\end{table}

\vspace{-0.12in}
\section{Experimental Results}
\vspace{-0.12in}
\label{sec:exp}
We now present qualitative and quantitative results on three
phantoms: a simulated phantom (\texttt{sim18}), a real-world
water bottle phantom (\texttt{Water}), and an actual baggage
phantom (\texttt{Clutter}).  We have implemented the
following algorithms and evaluated them on these three
phantoms: (i) \textbf{CDM-FBP}: CDM \cite{ying2006dual} with 16 CPU threads; (ii) \textbf{LM($m$)$\times$CG($n$)}: ADMM as in \cite{tracey2015stabilizing}; and (iii) \textbf{UDM($m$)-(P)CG($n$)}: proposed ADMM method with $m$ UDM iterations and $n$ (P)CG iterations.

Our Python implementations use the Astra toolbox
\cite{van2016fast} for GPU-based calculations of forward and
backward projections.  Additionally, as a key factor for
acceleration, we use Gpufit \cite{przybylski2017gpufit} to
parallelize the decomposition step.  All experiments are
carried out on a single computing cluster node with 16
cores, 20GB RAM and an Nvidia 1080Ti GPU.  Regarding
initialization, in our experiment we have chosen the CDM
output as the initial Compton estimate, since it is
generally stable.  For initial PE coefficients, we use a
scaled version of the Compton estimate, similar to what is
done in \cite{tracey2015stabilizing}.  For both Compton and
PE, we used $\lambda = 10^{-5}$ as the TV regularization
parameter and $\rho = 10^{-3}$ as the initial penalty
parameter.  Furthermore, for scale-invariant evaluation,
quantitative reconstruction quality is assessed with the
normalized \ltwo-distance between $\vx$ and ground truth
$\vx^*$:
\vspace{-0.08in}
\begin{equation}
\xi(\vx) = 20 \log_{10} \left (  \frac{\left \| \vx - \vx^* \right \|_2}{\left \| \vx^* \right \|_2}\right ).
\end{equation}

The first phantom, \texttt{sim18}, contains seven circular
regions of different materials.
% for
% evaluating the accuracy of the recovered Compton and PE
% coefficients.  
The reconstructed images, of size $512 \times
512$, are constructed from 720 angles,
each containg 725 parallel ray projections. The
first row in \fref{fig:recons} shows the reconstructions by
\textbf{CDM-FBP} and the proposed \textbf{UDM-PCG}
algorithm.  In \fref{fig:sim18_water_e_i}a, we display the
average computational time per iteration inside parentheses
in the inset box and plot $\xi(\vx)$ versus iteration to
compare the convergence rates.  In general, the ADMM
algorithm with the proposed splitting scheme results in
significantly shorter total execution time than 
\textbf{LM$\times$CG} to reach the same error.
% for the same fixed-point stopping criteria.

For results on the two real-world phantoms, the parallel
beam projection data for the two was collected on the
Imatron C300 CT scanner.  The X-ray source emits $1.8 \times
10^5$ and $1.7 \times 10^5$ photons per ray with two energy
spectra at 95keV and 130keV, respectively.  The high- and
the low-energy sinograms are subsampled by two, resulting in
360 angles for each and with 512 bins for each angle.  The reconstruction results for the
\texttt{Water} phantom are shown in the second row of
\fref{fig:recons}.  For quantitative evaluation, in
\fref{fig:sim18_water_e_i}b we plot $\xi(\vx)$ versus
iteration within the ROI -- the central circular region that
is occupied by distilled water.  
% For the computational
% effort, shown inside the parentheses in the inset box is the
% execution time per iteration.  
Our proposed
\textbf{UDM-PCG} algorithm is significantly more efficient
not only in terms of the number of the iterations needed,
but also in terms of the time per iteration as compared to
\textbf{LM$\times$CG}.  
% Additionally, the divergence of
% \textbf{ADMM-(P)CG5} indicates that `relaxed' updates are
% crucial for ADMM to be effective.

%% That brings us to the real-world baggage phantom
%% \texttt{Clutter}.  It is a particularly challenging phantom
%% for DECT because it contains metallic objects.  We use this
%% phantom in particular to illustrate the necessity of using a
%% statistical reconstruction approach.
The real-world baggage phantom \texttt{Clutter} is
particularly challenging because it contains metallic
objects, and illustrates well the need for a
statistical reconstruction.
As shown in the third
row of \fref{fig:recons}, while the \textbf{CDM-FBP} PE
reconstruction is completely overshadowed by streaking
artifacts, \textbf{UDM1-PCG1} is able to
recover object shapes to a reasonable degree.

\vspace{-0.12in}
\section{Conclusions and Future Work}
\vspace{-0.12in}
In this paper, we have proposed a new splitting scheme for
implementing ADMM to reconstruct Compton and PE coefficient
images using dual-energy projection data.  By separating the
the reconstruction and decomposition steps, the proposed
GPU-accelerated ADMM algorithm achieves a significant
speedup in time when compared to the prior state-of-the-art.
Future work will be aimed at generalization to 3D
reconstruction and improving initialization heuristics.

\bibliographystyle{IEEEbib}
\bibliography{refs}

\end{document}